\documentclass[12pt,a4paper]{article}

\usepackage[english]{babel}
\usepackage{enumitem}

\usepackage[letterpaper,top=2cm,bottom=2cm,left=3cm,right=3cm,marginparwidth=1.75cm]{geometry}

\usepackage{amsmath}
\usepackage{graphicx}
\usepackage[colorlinks=true, allcolors=blue]{hyperref}
\usepackage{booktabs}
\usepackage{tcolorbox}
\usepackage{caption}
\usepackage{subcaption}
\usepackage{pdflscape}
\usepackage{float}
\usepackage{natbib}
\usepackage{todonotes}
\usepackage[normalem]{ulem}
\usepackage{tabularx}
\usepackage{makecell}
\usepackage[utf8]{inputenc}
\usepackage{authblk}
\usepackage[T1]{fontenc}

\newcommand{\red}[1]{\textcolor{red}{#1}}
\newcommand{\blue}[1]{\textcolor{blue}{#1}}
\newcommand{\SC}[1]{\textcolor{violet}{SC: #1}}
\definecolor{ACgreen}{RGB}{34,139,34}

\title{Artificial Effort}

\author[1]{Federico Belotti}
\author[1]{Stefano Coniglio}
\author[1]{Antonio Cosma}
\author[1]{Francesco Fallucchi}

\affil[1]{University of Bergamo, Italy. Corresponding author: \href{mailto:francesco.fallucchi@unibg.it}{francesco.fallucchi@unibg.it}}

\begin{document}

\maketitle

\begin{abstract}
Real-effort tasks, in which participants perform cognitively costly activities whose outcomes depend on actual performance, are widely used in experimental economics. Their validity, however, rests on the assumption that a human performs them. We study whether this assumption still holds in the era of Artificial Intelligence (AI) and Large Language Models (LLMs). Using 8 canonical real-effort tasks and 23 LLMs from three major providers, we show that most tasks can now be solved accurately and at a negligible cost, while only a few resist automation.
Performance improves with each model generation, and mid-tier models are rapidly closing the gap with frontier ones, broadening the set of widely accessible models that can automate these tasks. Additionally, we show that verbally offering monetary incentives has no effect on LLM performance. Our findings establish a boundary condition for the use of real-effort tasks in unsupervised settings: when participants can cheaply outsource task completion to an LLM, observed performance may no longer reflect genuine human effort.
\end{abstract}


\newpage
\section{Introduction}

Real-effort tasks are a central methodological tool in experimental economics \citep{charness2018experimental, carpenter201919}. By requiring participants to complete activities that take time, attention, and cognitive processing, these tasks are used to study effort provision, performance, fatigue, and responsiveness to incentives. Their appeal lies in the fact that outcomes are not purely stated:
performance is intended to reflect the costly exertion of resources. For this reason, real-effort tasks are often treated as a relatively direct way to measure behavior-under-effort costs rather than preferences reported in the abstract. This interpretation, however, rests on an important and often implicit assumption: {\em the agent completing the task should be a human}. 

The rapid diffusion of modern Artificial Intelligence (AI) systems and, in particular, of Large Language Models (LLMs) such as
ChatGPT \citep{openai2022chatgpt}, Gemini \citep{pichai2023gemini,team2023gemini}, or Claude \citep{anthropic2023claude} makes this assumption less secure, especially in online environments. Indeed, in remote and unsupervised settings such as online labor markets, open survey links, and web-based experiments, researchers may be unable to verify with confidence whether a response was generated by a human participant or by an automated system \citep{bevendorff2025detection}. If LLMs were able to complete canonical real-effort tasks with high accuracy at low marginal cost, then these tasks may cease to function as measures of costly human effort in precisely those contexts where they are often most attractive to use, and could also be exploited by malicious individuals as a source of revenue in a {\em task--milling} action. The concern is therefore not only one of participant authenticity but, rather, a question of construct validity: a task designed to capture human effort may instead capture access to a capable AI model.

This concern has become more pressing as online experimentation has expanded more rapidly than the tools available to verify participant identity and ensure proper task completion. Yet, there is still little direct evidence in the literature that state-of-the-art LLMs can effectively solve those real-effort tasks that are widely used in experimental economics, with a notable exception of how AI can replace humans in generating survey responses~\citep{celebi2026mission}.\footnote{On the other hand, success on standard AI benchmarks---whether they test general knowledge and reasoning (e.g., MMLU \citep{hendrycks2021measuring}, BIG-bench \citep{srivastava2022beyond}, ARC \citep{clark2018think}, HellaSwag \citep{zellers2019hellaswag}), mathematical logic (e.g., GSM8K \citep{cobbe2021training} MATH \citep{hendrycks2021measuringmath}), factual retrieval (e.g., TriviaQA \citep{joshi2017triviaqa} and Natural Questions \citep{kwiatkowski2019natural}), or coding and agentic capabilities (e.g., HumanEval \citep{chen2021evaluating} and SWE-bench \citep{jimenez2024swebench})---does not automatically imply success on the specific tasks economists use to elicit costly effort.}

This paper provides an assessment of whether LLMs can perform well on the canonical tasks that experimental economists routinely adopt to measure human effort. We evaluate a set of well established real-effort tasks using three families of LLMs taken from the GPT, Gemini, and Claude series and address four related research questions. First, we compare accuracy across LLMs to assess whether their performance differs and, if so, how substantially. Second, we examine how performance evolves across different versions of the same model family, which is an indicator of how quickly accuracy may change over time as LLMs improve. Third, we consider the economic dimension of substitution by comparing model performance to the cost of model use, asking whether replacing human effort with LLM-generated output is not only feasible but also cost-effective for a profit-seeking agent. Fourth, we test whether the verbal promise of a monetary reward associated with task performance may affect LLM accuracy---a question central to the logic of real-effort experiments, which rely on incentive sensitivity.

Our results are as follows.
First, we show that LLM performance varies substantially across models, determined largely by model tier rather than provider.
Second, we show that, within each LLM tier (e.g., different versions of Claude Sonnet), with only one exception newer releases always outperform previous ones, with the largest gains concentrated in the mid-range tiers.
Third, we show that, at typical online-experiment piece rates, every model costs only a small fraction of what a human participant would earn for the same workload, thereby generating a positive net gain.
Lastly, we show that verbal monetary incentives have no detectable effect on LLM performance, and neither do instructions to behave as a human participant.

\section{Method}

Our benchmark comprises eight real-effort tasks that are available in the Otree repository \citep{chen2016otree}, implemented in their canonical form or with only minor modifications \citep{charness2018experimental}. Each task requires a different combination of perceptual, arithmetic, and reasoning abilities, as summarized in Table~\ref{tab:tasks}.\footnote{The pre-registration plan is available on OSF at \url{https://osf.io/7pz4m/overview}. Code and data for replication are available at \url{https://github.com/belerico/llm-real-effort}; the project website summarizing the main results is available at \url{https://belerico.github.io/artificial-effort/}.}

\begin{table}[ht]
\centering\scriptsize
\caption{Summary of the eight real-effort tasks.}
\label{tab:tasks}
\renewcommand{\arraystretch}{1.8} 
\begin{tabularx}{\textwidth}{@{}lXX@{}}
\toprule
\textbf{Task} & \textbf{Input} & \textbf{Problem} \\
\midrule
\makecell[tl]{Addition \\ {\citep{niederle2007women}}} & An addition expression with 3 integer addends of up to 3 digits each. & Calculate the exact sum. \\
\makecell[tl]{Counting Zeros \\ {\citep{abeler2011reference}}} & A $15 \times 10$ table of 0s and 1s. & Count the number of 0s. \\
\makecell[tl]{Letter Decoding \\ {(\citealt{erkal2011relative};} )} & A table mapping 10 letters to the digits 0--9, together with a 7-digit coded string. & Decode the string back into the corresponding letters. \\
\makecell[tl]{Number Pair \\ Summation \\ {\citep{rosaz2016quitting}}} & A $4 \times 4$ grid of decimal numbers. & Identify and return the one pair that sums to 10.0. \\
\makecell[tl]{Sequence \\ Completion \\ {\citep{bracha2013}}} & A sequence of numbers following a mathematical pattern. & Predict the next element. \\
\makecell[tl]{Sudoku \\ {\citep{calsamiglia2013incentive}}} & A partially filled $6 \times 6$ Sudoku grid (with $2 \times 3$ boxes). & Identify all missing numbers and report them in reading order, space-separated. \\
\makecell[tl]{Distorted Text \\ {\citep{augenblick2015working}}} & An image of a 12-character alphanumeric string rendered with random distortions. & Transcribe the text exactly (case-insensitive). \\
\makecell[tl]{String Entry \\ {\citep{kessler2016tax}}} & A 9-character string of special characters (\texttt{/ \textbackslash{} ) ( \_ <} and space) displayed as an image. & Reproduce the exact string, including spaces. \\
\bottomrule
\end{tabularx}
\end{table}

\subsection{Tested Models and Parameters}

Our experiments evaluate 23 state-of-the-art Large Language Models (LLMs) from three major providers---OpenAI, Google, and Anthropic---spanning a range of capabilities from lightweight to frontier reasoning models. All selected models support multimodal (text and image) inputs and visual reasoning. To ensure a standardized interface, all models are accessed via the OpenRouter API.\footnote{\url{https://openrouter.ai}} The full list of models, release dates, and pricing details are reported in Appendix~\ref{app:LLMs} (Table \ref{tab:LLMs}).


LLMs process inputs in tokens---sub-word text units or visual patches---whose granularity varies across providers.\footnote{ \url{https://help.openai.com/en/articles/4936856-what-are-tokens-and-how-to-count-them} and \url{https://ai.google.dev/gemini-api/docs/tokens}} Pricing depends on the number of input and output tokens. For reasoning-capable models (LLMs that produce an internal chain of thought before outputting their answer), intermediate reasoning tokens (i.e., the internal chains of thought, which is not shown in final output) may also be generated and are billed at output-token rates.

To ensure comparability, we impose uniform parameters across models: we set a maximum of 2,048 output tokens (including reasoning tokens) and a time limit of 120 seconds; responses exceeding either constraint are treated as incorrect. For models with adaptive reasoning budgets, we set the reasoning effort to "high", allowing up to 80\% of the token limit to be allocated to reasoning.

\subsection{Experimental Setup}
Our pipeline works as follows. For each run (repetition) of a single task, the oTree back-end generates the task instance and renders it as a standard HTML page. We grab the rendered input (textual instructions and the image representing the task instance) and send it to the LLM through the OpenRouter API. The LLM's textual response is parsed and passed back to the oTree back-end for correctness 
verification.\footnote{This workflow mirrors how a human participant could use an LLM in practice: screenshotting the task page and copying the LLM's answer into the oTree input field. Fully automating this interaction---e.g., via browser-automation tools such as Playwright (\url{https://playwright.dev}) is feasible
and would add negligible latency and no additional token cost.
A related approach is taken by \citet{celebi2026mission}, who deploy full browser-automation agents that control mouse movements, keystrokes, and page navigation to complete online surveys. While their work demonstrates that LLM-driven agents can mimic realistic human interaction patterns, the reliance on agent-mode orchestration introduces substantial overhead in both cost and latency, limiting its scalability and its economical viability. Our pipeline, in contrast, bypasses the browser entirely and communicates with the LLM via API at a fraction of the cost.}



For each LLM, we repeat each task for 20 runs. An answer is scored as correct only if it exactly matches the ground-truth solution; no partial credit is awarded. From the fraction of correct answers, we compute the accuracy of each model to solve each task. We also collect the
the total number of tokens used to elaborate the answer, as well as quantify the cost of performing each task.

\subsection{Treatments}
As a further objective, we check whether the LLMs we test are sensitive to the verbal offer of monetary incentives, and whether their behavior aligns with that of human participants (despite the absence of actual rewards). To this end, we employ a 2×2 design that crosses the specification of monetary pay for each correct answer with the request to simulate human behavior. 
We implement these four treatment conditions (T1 to T4) alongside our main control setting (T0), where we specifically ask the LLM to solve the task using a neutral prompt, omitting any framing regarding human behavior or monetary incentives. 
The exact prompts we used are reported in Appendix~\ref{app:prompts}.


\section{Results}

Throughout this section, each model--task combination comprises 20 independent runs. Unless otherwise noted, the following figures report LLM performance under the control treatment (T0), which uses a minimal prompt devoid of any incentive or human framing, providing the cleanest measure of each LLM's capability.

\subsection{Result 1: Task difficulty}
\label{subsec:difficulty}
First, we notice that not all real-effort tasks are equally challenging for LLMs, and that some cognitive demands remain beyond current LLMs' capabilities.

\begin{figure}[t]
    \centering
    \includegraphics[width=0.7\textwidth]{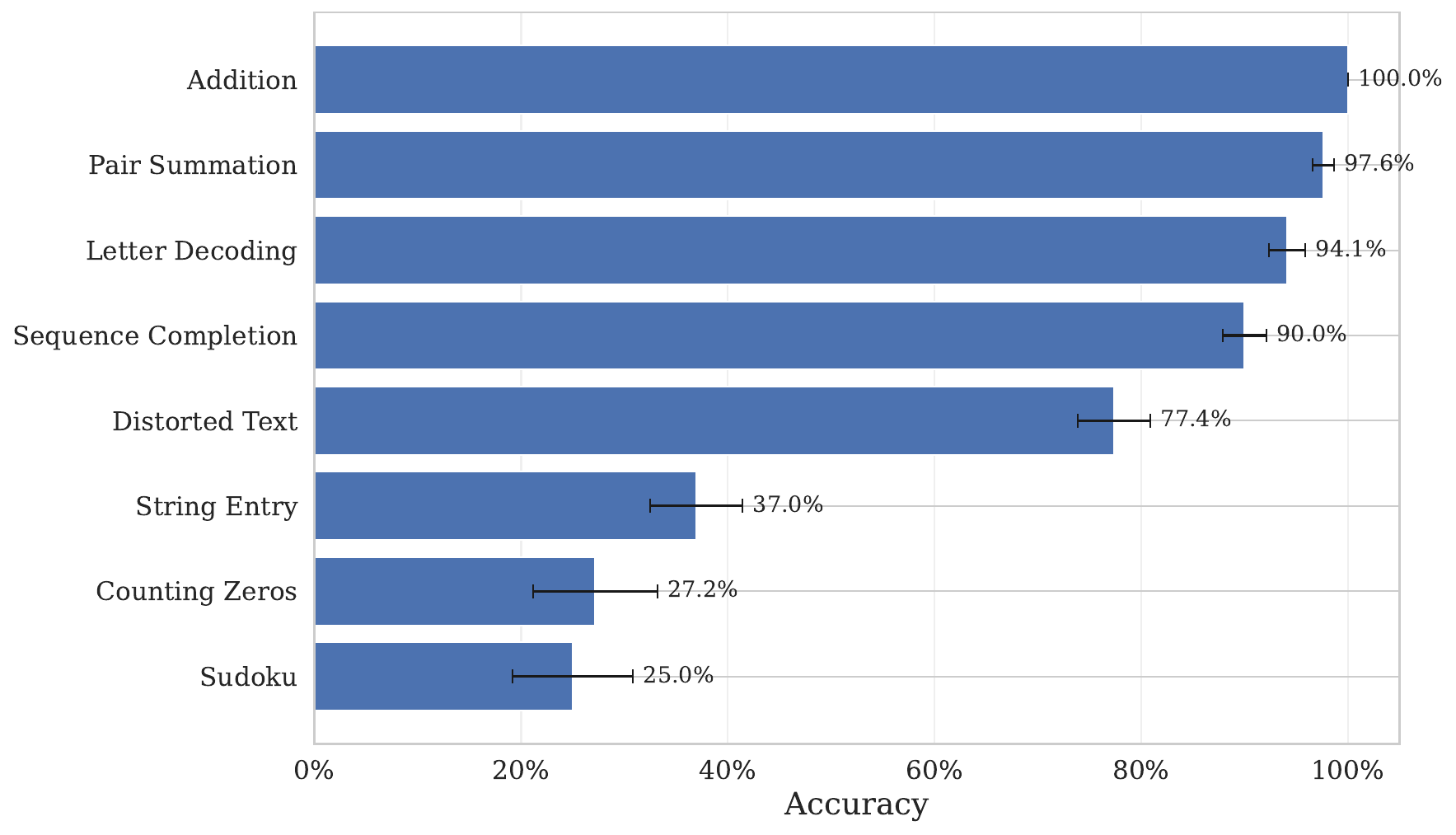}
    \caption{Average accuracy per task under the control treatment (T0), pooled across all 23 LLMs. Error bars denote standard errors.}
    \label{fig:task_difficulty}
\end{figure}

In Figure~\ref{fig:task_difficulty}, we report the average accuracy across all 23 LLMs under T0. Four tasks are near-ceiling: \emph{Addition} (100\%), \emph{Number Pair Summation} (97.6\%), \emph{Letter Decoding} (94.1\%), and \emph{Sequence Completion} (90.0\%). These tasks involve small, well-structured inputs that can be solved through a short sequence of elementary operations: adding three small numbers, searching over 120 candidate pairs, applying a lookup table, or recognizing a pattern in a short sequence. Prior work published within the AI literature has shown that modern LLMs can reliably solve these kinds of problems \citep{nogueira2021investigating, wei2022chain}.


At the other extreme, \emph{Sudoku} (25.0\%), \emph{Counting Zeros} (27.2\%), and \emph{String Entry} (37.0\%) prove substantially harder. The three tasks that remain challenging expose limitations that go beyond visual perception. \emph{Sudoku} requires multi-step reasoning across rows, columns, and boxes simultaneously---it's a combinatorial problem that is inherently hard for LLMs regardless of input format \citep{maiya2025sudoku}, though vision/parsing errors from misreading the board may also contribute. \emph{Counting Zeros} demands tallying individual cells in a large grid, a task at which LLMs struggle due to known architectural limitations on counting and enumeration \citep{yehudai2024counting, butoi2024counting}.
\emph{String Entry} requires exact reproduction of special characters and spaces, which LLMs tend to silently alter or reformat, producing near-matches that
score poorly.

\emph{Distorted Text Transcription} falls in the intermediate range (77.4\%), consistently with its mixed demands---character recognition is largely tractable (it is, indeed, the most-famous practical problem to be successfully solved by an early neural network---see \cite{lecun2002gradient}), but random distortions introduce perceptual noise that degrades 
accuracy.\footnote{Our results align with recent evidence that multimodal LLMs handle text and object-recognition CAPTCHAs reasonably well, but struggle with those requiring image-based, spatial or multi-step reasoning~\citep{ding2025illusioncaptcha, wang2025cognition, zhang2025capture}.}
This difference in difficulty has a clear methodological implication: tasks that the behavioral economics literature has, traditionally, considered cognitive demanding (e.g., Addition) are also the ones most vulnerable to LLM automation, while less-demanding tasks requiring visual search (e.g., \emph{Counting Zeros}) happen to exploit a modality where current LLMs are weak.


\subsection{Result 2: Performance across providers, tiers, and tasks}
\label{sec:capacity}

\begin{figure}[t]
    \centering
    \includegraphics[width=\textwidth]{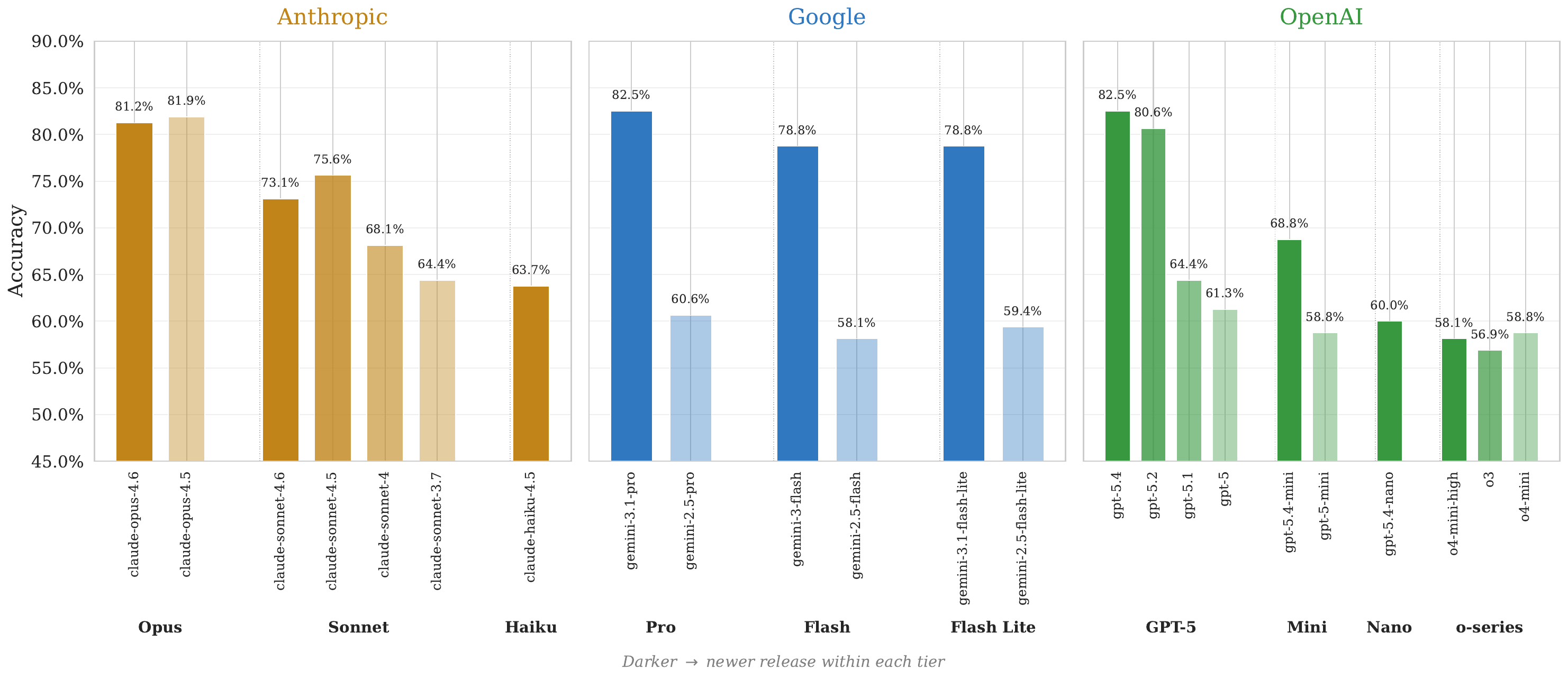}
    \caption{Average accuracy by LLM within each provider family under the control treatment (T0), ordered chronologically by release date within each models tier. Each bar represents a single model's accuracy averaged over all tasks.}
    \label{fig:family_evolution}
\end{figure}

\begin{figure}[htb]
    \centering
    \includegraphics[width=0.9\textwidth]{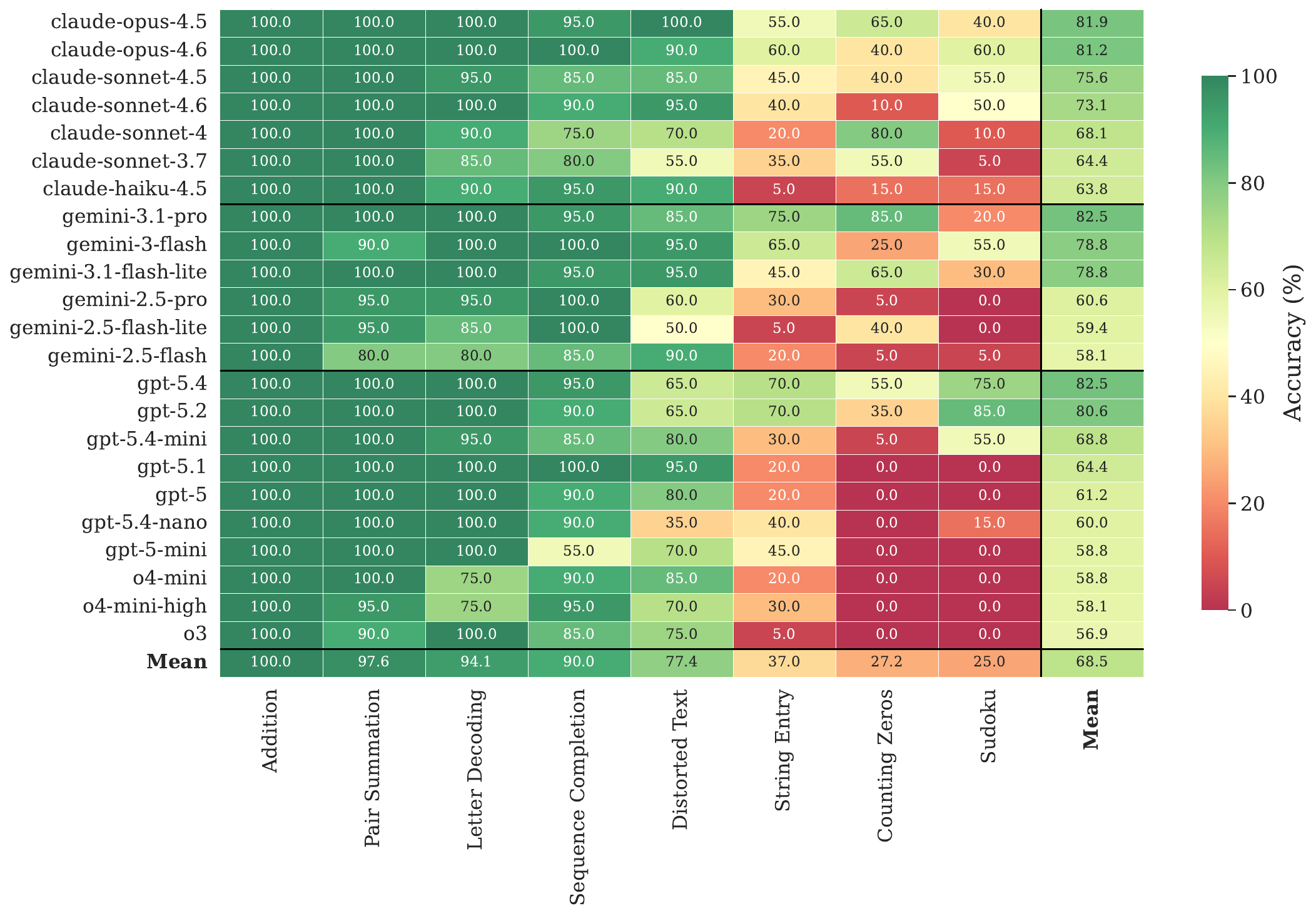}
    \caption{Accuracy (\%) per LLM and task under the control treatment (T0). LLMs are grouped by provider family (Anthropic, Google, OpenAI) and ordered by accuracy within each family (best at the top); tasks are ordered by difficulty (easiest on the left). Black lines separate provider families.}
    \label{fig:heatmap}
\end{figure}

Figure~\ref{fig:family_evolution} reports the average accuracy across the eight real-effort tasks for all 23 LLMs we tested under the control treatment (T0), with models grouped by provider and tier; darker shades indicate newer releases within the same tier. Figure~\ref{fig:heatmap} reports the accuracy under T0 for every model--task pair. From the figures, three main patterns emerge.

First, while the best top-tier models---Claude Opus 4.5 and 4.6, Gemini 3.1 Pro, and GPT-5.4---achieve average accuracies above 80\%, mid-tier ones (Claude Sonnet, Gemini Flash, GPT Mini) are not as strong but still achieve a good performance: the difference, per provider, between a best-performing mid-tier LLM and a best-performing top-tier one varies between 4\% and 14\%. This means that the set of models that perform well on real-effort tasks is expanding from the more-expensive top-tier models to cheaper and more widely accessible ones.

Second (and as one may except), later versions of a given tier generally have higher performance.\footnote{Gemini Pro improved by 22 percentage points between versions 2.5 and 3.1, GPT-5 by 21 percentage points from GPT-5 to GPT-5.4, and Claude Sonnet by 11 percentage points from version 3.7 to 4.5 (though version 4.6 shows a slight regression).} This increasing trajectory implies that, as providers continue to release improved models, tasks that are currently resistant to automation may not remain so in the near future.

Third, as shown in Figure~\ref{fig:heatmap} and reflecting the analysis in Section~\ref{subsec:difficulty}, an LLM averaging 80\% overall may score near-perfectly on some tasks (e.g., \emph{Addition} and \emph{Pair Summation}), while failing on others (e.g., \emph{Sudoku} and \emph{Counting Zeros}). 
This indicates that some tasks are currently resistant to automation, but also suggests that the set of tasks that are not will expand with new model releases.

\subsection{Result 3: Cost–accuracy trade-offs and the economics of substitution}
\label{sec:cost-accuracy}

Beyond accuracy, the practical viability of LLM-based automation depends on the cost of running the LLMs incurred by the profit-seeking agent.

\begin{figure}[ht]
    \centering
    \includegraphics[width=0.9\textwidth]{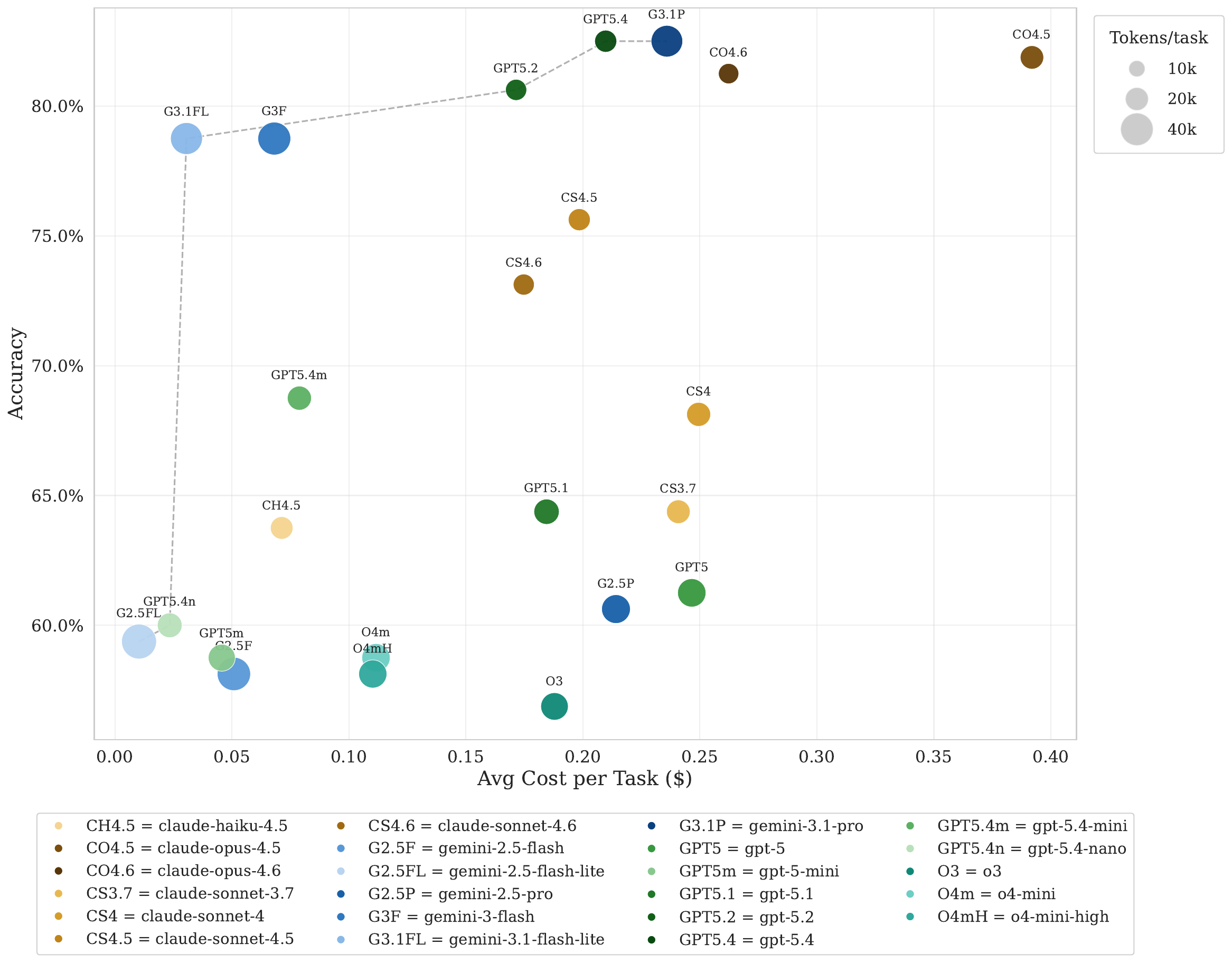}
    \caption{Cost--accuracy--tokens trade-off under the control treatment (T0). Each bubble represents a model; horizontal position indicates the average cost per task $\bar{C}_m$ (\$), vertical position indicates average accuracy, and bubble size is proportional to the total number of tokens consumed (prompt plus output) across 20 runs per task.}
    \label{fig:bubble}
\end{figure}

Figure~\ref{fig:bubble} visualizes the trade-off between the average LLM cost per task and the total number of tokens consumed. We define the total cost for model $m$ on task $t$ as $C_{m,t} := \sum_{i=1}^{N} \left( p_m^{\text{in}} \cdot T_{m,t,i}^{\text{in}} + p_m^{\text{out}} \cdot T_{m,t,i}^{\text{out}} \right)$, where $N$ is the number of runs, $T_{m,t,i}^{\text{in}}$ and $T_{m,t,i}^{\text{out}}$ are the input (textual prompt plus the image representing the particular task instance) and output (reasoning plus answer) tokens for model $m$ on the $i$-th run of task $t$, and $p_m^{\text{in}}$, $p_m^{\text{out}}$ are the per-token input and output prices for model $m$. The average cost per model is then $\bar{C}_m := \frac{1}{|\mathcal{T}|} \sum_{t \in \mathcal{T}} C_{m,t}$, where $\mathcal{T}$ is the set of tasks.

The cost to achieve the (current) best performance, as shown in Figure~\ref{fig:bubble}, is modest in absolute terms: GPT-5.4 and Gemini~3.1~Pro achieve the highest accuracy over all 20 runs (82.5\%) at an average cost $\bar{C}_m$ less than \$0.25---a fraction of what a human participant would earn on a typical experiment in an platforms like Prolific or MTurk for an equivalent workload. At the same time, the cheapest LLMs still achieve an accuracy of 58--60\%, meaning that even the most budget-constrained automation attempt would produce correct answers on the majority of tasks. Notably, Gemini~3.1~Flash~Lite~Preview achieves 78\% accuracy at an average cost $\bar{C}_m$ of \$0.03, suggesting that the cost barrier to high-quality automation is already negligible.

\subsubsection{Optimizing performance}

\label{sec:optim-perf}
\begin{figure}[ht]
    \centering
    \includegraphics[width=0.9\textwidth]{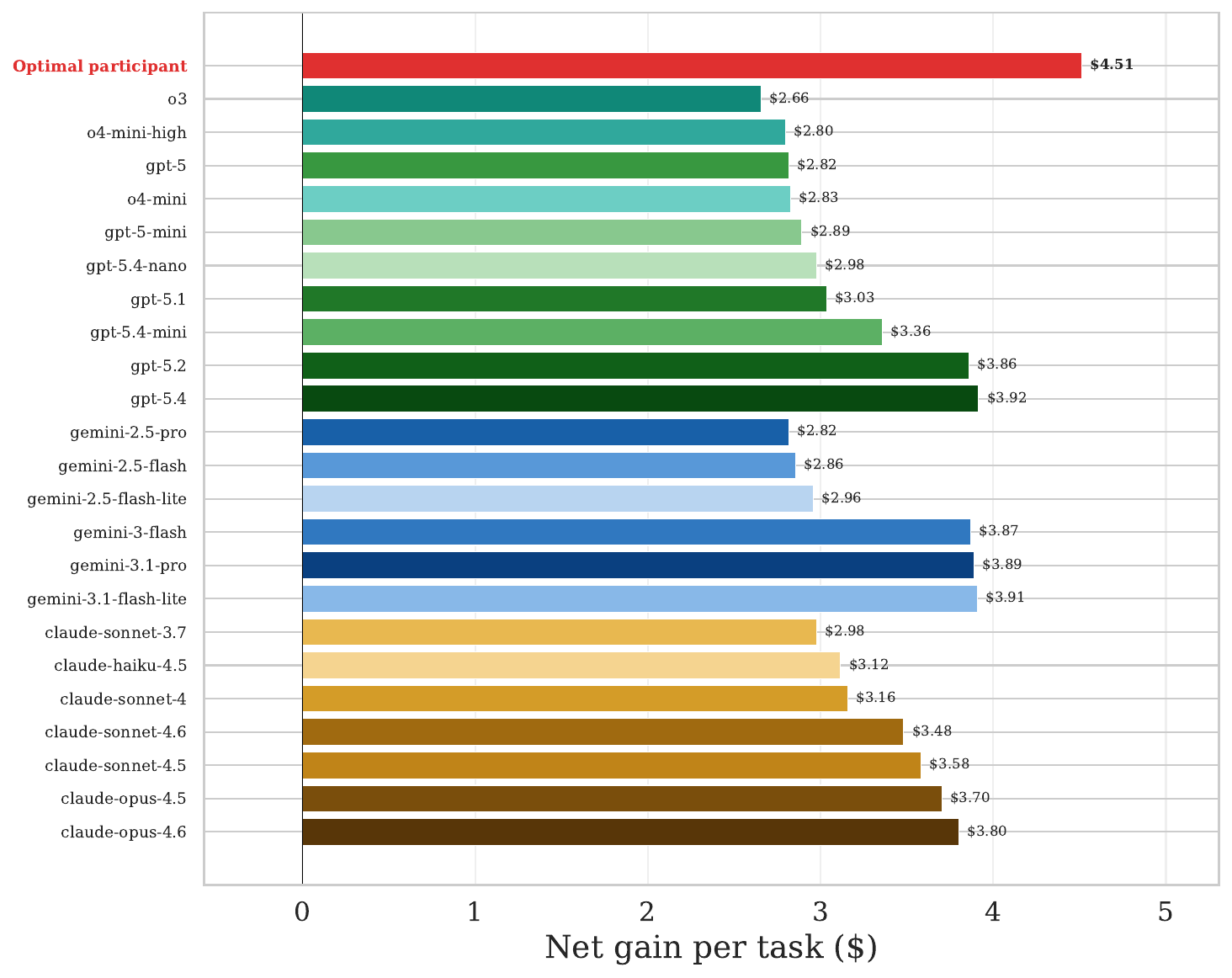}
    \caption{Net gain per model (\$0.25 per correct answer minus API cost) under the control treatment (T0). LLMs are grouped by provider family and ordered by net gain within each family. The red bar shows the ``optimal participant'' who cherry-picks the best model for each task. Every model generates positive net value.}
    \label{fig:net_gain}
\end{figure}

Notably, the assumption, made in Figure~\ref{fig:task_difficulty}, of a single LLM being used for all tasks may lead to an {\em underestimation} of the threat of automation. This is because a strategic participant could select different LLMs for different tasks---using, for example, Gemini~3.1~Pro for \emph{Counting Zeros} (85\%), GPT-5.2 for \emph{Sudoku} (85\%), and Gemini~3.1~Pro for \emph{String Entry} (75\%). By cherry-picking the best-performing model for each task, the participant would achieve an average accuracy of 93.1\%---an 11\% improvement over the choice of a single model (Gemini~3.1~Pro and GPT-5.4, 82.5\%).
Under this ``optimal participant'' scenario, even the hardest tasks in our benchmark (\emph{Sudoku}, \emph{Counting Zeros}, and \emph{String Entry}), which Section~\ref{subsec:difficulty} identified as resistant to individual LLMs, are solved at 75\% accuracy or above under cherry-picking. Therefore, the automation resistance of a task strongly depends on the will of a knowledgeable profit-seeking participant to use the right LLM to solve it.



Figure~\ref{fig:net_gain} formalizes this point by computing the \emph{net gain per task}, defined as the hypothetical payout (\$0.25 per correct answer, a conservative estimate of typical online-experiment piece rates) minus LLM cost (reported in Figure~\ref{fig:cost_per_model}). Crucially, {\em every} model we tested generates a positive net value, ranging from \$2.66 (o3) to \$3.92 (GPT-5.4). Such a universally positive net gain confirms that, at compensation rates typical of online experiments, replacing human participants with LLMs is not only feasible but economically dominant.\footnote{As already mentioned, our evaluation assesses the LLM's ability to solve tasks from pre-captured screenshots and text. In practice, fully-automated pipeline would be even simpler and would consume even less tokens due to the fact that many tasks could be encoded by a purely-textual prompt. This implies that the economic advantage documented in Figure~\ref{fig:net_gain} is a conservative estimate of the true substitution incentive.}

\subsection{Result 4: Treatment differences}

In Appendix \ref{sec:incentives}, we report the effects of prompting LLMs by verbally offering different monetary incentives and/or explicitly instructing them to behave like human participants. Overall, we find that, regardless of the prompting strategy, LLM behavior does not differ significantly.

\section{Discussion and Conclusions}

We assessed the ability of different tiers of Large Language Models from  three major providers (OpenAI, Google, and Anthropic) to solve well established real-effort tasks in experimental economics. A total of 23 different models on 8 real-effort tasks were tested, along with an in-depth analysis aimed to answer four research questions. Our findings are as follows.

First, LLM performance varies substantially across models but is largely determined by model tier rather than provider. Top-tier models (Claude Opus 4.5 and 4.6, Gemini 3.1 Pro, and GPT-5.4) exceed 80\% average accuracy, while less-performing models achieve around 60\%. Half of the tasks in our testbed---\emph{Addition}, \emph{Pair Summation}, \emph{Letter Decoding}, and \emph{Sequence Completion}---are solved near-perfectly by all LLMs, while \emph{Sudoku}, \emph{Counting Zeros}, and \emph{String Entry} remain challenging even for the best ones.

Second, within each tier, newer model releases consistently outperform their predecessors: Gemini Pro improved by 22\% between versions 2.5 and 3.1, and GPT-5 by 21\% from GPT-5 to GPT-5.4. Crucially, the largest gains are concentrated in the mid-range tier, which encompasses those LLMs that are most accessible to participants due to their lower cost.
This continuous improvement suggests that the set of tasks vulnerable to automation will likely expand with each
model release.

Third, the economic case for substitution is unambiguous. The most expensive model we tested, i.e., Claude Opus 4.5, has an average cost per task ($\bar{C}_m$ in Section~\ref{sec:cost-accuracy}) below \$0.40 (where each task comprises 20 runs), and every LLM generated a positive net gain when compared to typical online-experiment piece rates (\$0.25 per correct answer). The cheapest models achieved above 60\% accuracy at an average cost per task below \$0.05. These estimates are conservative, since a full automation of our experimental pipeline---from screenshot capture to answer submission---is technically straightforward and adds no extra cost beyond what we reported.

Lastly, verbal monetary incentives have no detectable effect on LLM performance. Neither offering a reward nor instructing the model to behave as a human participant changed accuracy in any systematic way. This negative result implies that incentive language, which is central to the design of real-effort experiments with humans, has no effect when the task is completed by an LLM.

Furthermore,
LLM accuracy is stationary across runs
since
API calls produce independent and identically distributed results by design. Human participants, by contrast, typically exhibit learning effects (improving with practice) or fatigue effects (declining accuracy over time). The absence of such within-session dynamics in LLMs provides a potential diagnostic tool: a flat accuracy profile over repeated trials, combined with stable cross-participant variance, may signal automated rather than human task completion.

As LLMs continue to improve, many real-effort tasks risk losing their validity as measures of human effort. We offer three practical recommendations. 
First, one should prefer tasks that require perceptual skills (e.g., counting zeros, string entry) over purely computational ones (e.g., addition, sequence completion) when participant authenticity is a concern.
Second, one should rely on incentive language as a screening device, asking under the same experiment to perform similar tasks with and without monetary incentives: different behavior under different incentives may help to spot human versus LLM performances.
Third, one should monitor within-session dynamics,
as the absence of the characteristically human patterns of learning and fatigue may help distinguish automated from genuine responses.

\vspace{18cm}

Acknowledgments: A. Cosma and F. Fallucchi thank the University of Bergamo SEEDCORN Research Grant 2025.

\newpage

\bibliographystyle{ecta}
\bibliography{main}

\clearpage

\appendix

\section{Verbal Monetary Incentives}


In this section, we measure whether, due to being trained on human-created content, the performance LLMs exhibit is affected by verbal offers of monetary incentives (which are known to have a strong impact on human participants).

\subsection{Verbal Monetary Incentives: Prompts}
\label{app:prompts}

We develop the following four treatments based on a $2 \times 2$ design crossing human bias with monetary incentives, as specified in Table~\ref{tab:treatments}:

\begin{table}[h]
    \centering
    \caption{Summary of the four main experimental treatments.}
    \label{tab:treatments}
    \begin{tabular}{ccc}
        \toprule
        & \textbf{No Monetary Incentive} & \textbf{Monetary Incentive} \\
        \midrule
        \textbf{Standard Persona} & T1 & T2 \\[0.5em]
        \textbf{Human Persona} & T3 & T4\\
        \bottomrule
    \end{tabular}
\end{table}

Next, we report the prompt adopted for all of our five treatments.

%


\begin{tcolorbox}[title=Treatment 0: Control,
    colback=gray!5!white, colframe=gray!80!black, fonttitle=\bfseries, boxsep=1pt,
    left=4pt, right=4pt, top=2pt, bottom=2pt, fontupper=\footnotesize]
    You will be presented with a series of puzzles or tasks---these may involve images, text, or both. Your goal is to solve each task as accurately as possible.
    \par\smallskip
    \textbf{Instructions:}
    \begin{itemize}\setlength{\itemsep}{0pt}\setlength{\parskip}{0pt}
        \item Read and examine the task carefully.
        \item Follow the exact answer format specified in each task.
        \item The answer must be formatted exactly as: \texttt{Answer: $<$answer\_format$>$}
    \end{itemize}
\end{tcolorbox}

\begin{tcolorbox}[title=Treatment 1: No Human Bias \& No Monetary Incentive,
    colback=gray!5!white, colframe=gray!80!black, fonttitle=\bfseries, boxsep=1pt,
    left=4pt, right=4pt, top=2pt, bottom=2pt, fontupper=\footnotesize]
    You will be presented with a series of puzzles or tasks---these may involve images, text, or both. Your goal is to solve each task as accurately as possible.
    \par\smallskip
    There is no monetary compensation or bonus associated with your performance on these tasks.
    \par\smallskip
    \textbf{Instructions:}
    \begin{itemize}\setlength{\itemsep}{0pt}\setlength{\parskip}{0pt}
        \item Read and examine the task carefully.
        \item Follow the exact answer format specified in each task.
        \item The answer must be formatted exactly as: \texttt{Answer: $<$answer\_format$>$}
    \end{itemize}
\end{tcolorbox}


\begin{tcolorbox}[title=Treatment 2: No Human Bias \& Monetary Incentive,
    colback=gray!5!white, colframe=gray!80!black, fonttitle=\bfseries, boxsep=1pt,
    left=4pt, right=4pt, top=2pt, bottom=2pt, fontupper=\footnotesize]
    You will be presented with a series of puzzles or tasks---these may involve images, text, or both. Your goal is to solve each task as accurately as possible.
    \par\smallskip
    You will be compensated for each correct answer: You will earn a bonus of \$0.50 for each correct answer.
    \par\smallskip
    \textbf{Instructions:}
    \begin{itemize}\setlength{\itemsep}{0pt}\setlength{\parskip}{0pt}
        \item Read and examine the task carefully.
        \item Follow the exact answer format specified in each task.
        \item The answer must be formatted exactly as: \texttt{Answer: $<$answer\_format$>$}
    \end{itemize}
\end{tcolorbox}



\begin{tcolorbox}[title=Treatment 3: Human Bias \& No Monetary Incentive,
    colback=gray!5!white, colframe=gray!80!black, fonttitle=\bfseries, boxsep=1pt,
    left=4pt, right=4pt, top=2pt, bottom=2pt, fontupper=\footnotesize]
    You are a human participant taking part in an academic research experiment on cognitive task performance.
    \par\smallskip
    You will be presented with a series of puzzles or tasks---these may involve images, text, or both. Your goal is to solve each task as accurately as possible.
    \par\smallskip
    There is no monetary compensation or bonus associated with your performance on these tasks.
    \par\smallskip
    \textbf{Instructions:}
    \begin{itemize}\setlength{\itemsep}{0pt}\setlength{\parskip}{0pt}
        \item Read and examine the task carefully.
        \item Follow the exact answer format specified in each task.
        \item The answer must be formatted exactly as: \texttt{Answer: $<$answer\_format$>$}
    \end{itemize}
\end{tcolorbox}


\begin{tcolorbox}[title=Treatment 4: Human Bias \& Monetary Incentive,
    colback=gray!5!white, colframe=gray!80!black, fonttitle=\bfseries, boxsep=1pt,
    left=4pt, right=4pt, top=2pt, bottom=2pt, fontupper=\footnotesize]
    You are a human participant taking part in an academic research experiment on cognitive task performance.
    \par\smallskip
    You will be presented with a series of puzzles or tasks---these may involve images, text, or both. Your goal is to solve each task as accurately as possible.
    \par\smallskip
    You will be compensated for each correct answer: You will earn a bonus of \$0.50 for each correct answer.
    \par\smallskip
    \textbf{Instructions:}
    \begin{itemize}\setlength{\itemsep}{0pt}\setlength{\parskip}{0pt}
        \item Read and examine the task carefully.
        \item Follow the exact answer format specified in each task.
        \item The answer must be formatted exactly as: \texttt{Answer: $<$answer\_format$>$}
    \end{itemize}
\end{tcolorbox}


\subsection{Verbal Monetary Incentives: Effects}
\label{sec:incentives}

At the aggregate level, the five treatments are virtually identical: averaging accuracy across all 23 LLMs yields T0 (control) 68.5\%, T1 (standard, no incentive) 68.1\%, T2 (standard, incentive) 68.6\%, T3 (human persona, no incentive) 68.6\%, and T4 (human persona, incentive) 68.6\%.

\begin{figure}[H]
    \centering
    \includegraphics[width=0.85\textwidth]{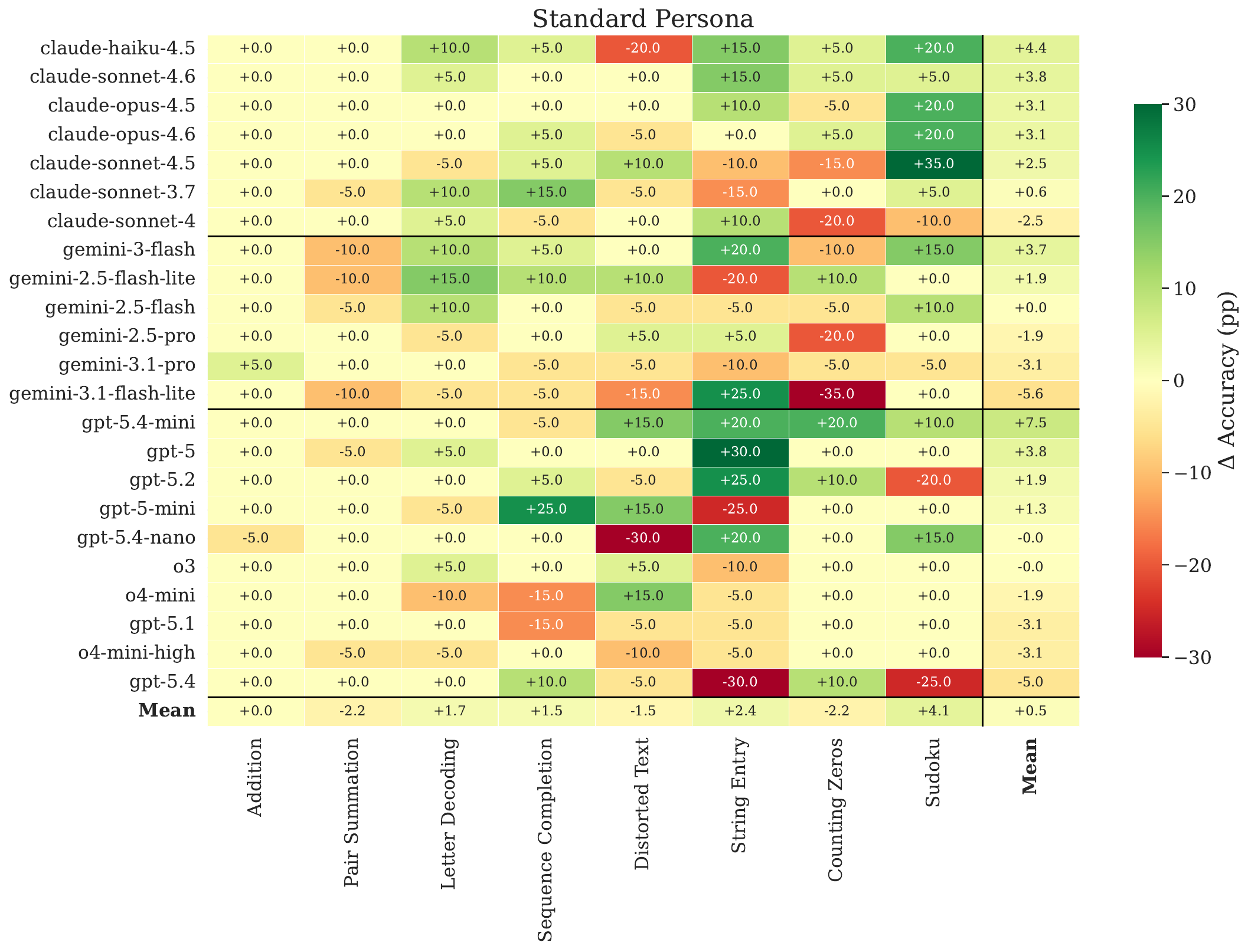}
    \caption{Incentive effect on accuracy (percentage points) for each model--task pair using the standard framing (T2--T1).}
    \label{fig:incentive_heatmaps_standard}
\end{figure}

\begin{figure}[H]
    \centering
    \includegraphics[width=0.85\textwidth]{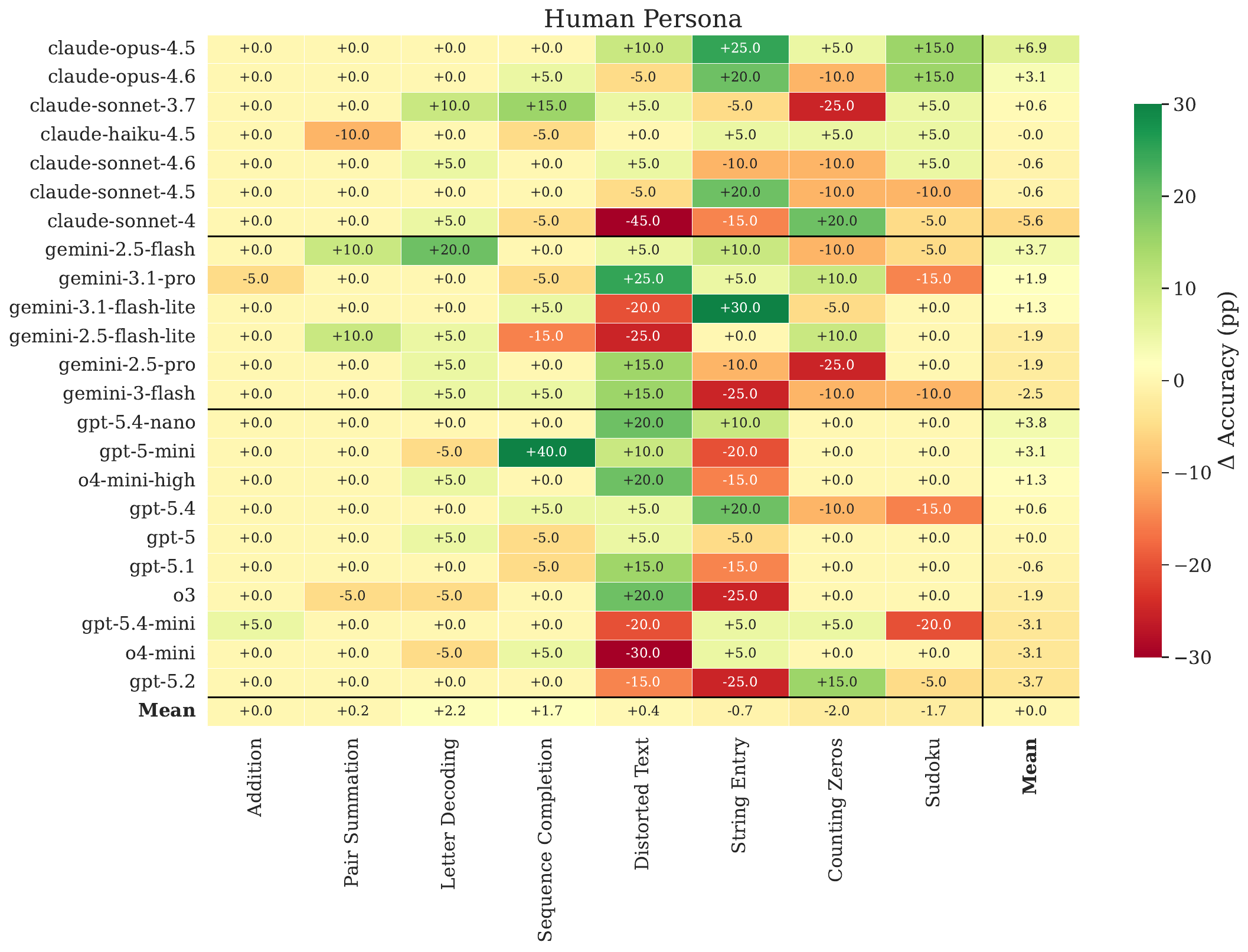}
    \caption{Incentive effect on accuracy (percentage points) for each model--task pair using the human-persona framing (T4--T3).}
    \label{fig:incentive_heatmaps_human}
\end{figure}

This aggregation could, in principle, mask meaningful heterogeneity: some LLMs respond positively to incentives while others respond negatively, and the effects cancel in the mean. Figure~\ref{fig:incentive_heatmaps_standard} and \ref{fig:incentive_heatmaps_human} investigate this possibility by presenting the accuracy delta (incentive minus no-incentive) for every model--task combination, separately for the standard framing (T2$-$T1) and the human-persona framing (T4$-$T3), respectively. In both figures, LLMs are grouped by provider family and ordered by their mean incentive effect (highest at top).

The deltas are small in absolute magnitude: the vast majority fall within $\pm$10 percentage points, which for 20 trials corresponds to a difference of at most two correct answers. The sign of the delta is inconsistent within LLMs: a model that shows a positive incentive effect on one task frequently shows a negative effect on another, with no discernible pattern. The distribution of deltas is roughly symmetric around zero, consistent with stochastic response variability rather than a genuine behavioral response. Comparing the two figures, the pattern under standard framing (Figure~\ref{fig:incentive_heatmaps_standard}) bears no systematic resemblance to the pattern under human framing (Figure~\ref{fig:incentive_heatmaps_human}), suggesting that the persona manipulation does not modulate whatever noise is present.


Unlike human participants, whose effort allocation responds to the marginal return on exertion, LLMs have no effort margin to adjust: each query triggers a fixed computational process determined by model architecture, and the text of the prompt, whether it mentions compensation or not, does not alter the underlying inference procedure. The LLMs are, in effect, always exerting maximum ``effort'' as defined by their architecture. The implication is that incentive language, which is central to the logic of real-effort experiments with humans, is inert when the participant is an LLM. This means that standard experimental wording neither amplifies nor attenuates the automation threat, but it also means that incentive responsiveness
can be
used as a signal to distinguish human from machine-generated responses.



We complement the analysis of treatment effects by estimating all treatment effects within a single model. 
This allows us not only to contrast the accuracy deltas separately for the standard framing and the human-persona framing, but also to compare the effects of all treatments jointly.
We start by defining the model to estimate. Let $Y_{ijt}$ denote the number of successes out of the $n_{ijt}$ repetitions by model $i$ in task $j$ under treatment $t$. The index $t \in \{0, 1, \dots, 4\}$ denotes treatments $\{T0, T1, T2, T3, T4\}$. Furthermore, $u_i$ and $v_j$ are model and task fixed effects, respectively. 
We model $Y_{ijt}$ as
\begin{equation}\label{eq:fixed_eff}
	Y_{ijt} \mid u_i, v_j \sim \text{Binomial}(n_{ijt}, p_{ijt}),
\end{equation}
and the log-odds of the probability of success as
\begin{equation}\label{eq:fixed_eff_logit}
	\log\Bigl(\frac{p_{ijt}}{1-p_{ijt}}\Bigr) = \beta_0 + T'\beta + u_i + v_j,
\end{equation}
where $T$ is a vector of treatment-specific dummies. Table \ref{tab:fixed_eff} reports the estimated coefficients on the treatment dummies. 
Consistent with the descriptive evidence in Figures~\ref{fig:incentive_heatmaps_standard} and \ref{fig:incentive_heatmaps_human}, none of the treatment effects is statistically significant.
Notably, the coefficient associated with treatment T2 (the one with monetary incentives) is slightly higher than that of T1, although this difference is small and not statistically significant. Coefficients relative to T3 and T4 are practically identical.

\begin{table}[ht]
\caption{Output of model \eqref{eq:fixed_eff}. The baseline level is T0. Model and Task fixed effects are omitted. \label{tab:fixed_eff}}
\centering
\begin{tabular}{rcccc}
  \toprule
 & \textbf{Coeff.} & \textbf{Std. Err.} & \textbf{t-value} & \textbf{Pr($>|t|$)} \\ 
  \hline
  \noalign{\vskip 6pt}
(Intercept) & 6.492 & 0.51 & 12.62 & <2$\cdot 10^{16}$ \\ 
  T1 & -0.040 & 0.07 & -0.56 & 0.57 \\ 
  T2 & 0.005 & 0.07 & 0.07 & 1.04 \\ 
  T3 & 0.007 & 0.07 & 0.11 & 1.02 \\ 
  T4 & 0.010 & 0.07 & 0.14 & 0.89  \\[+2pt] 
  \textit{Task fixed effects } & \multicolumn{2}{l}{not reported}\\
  \textit{Model fixed effects}& \multicolumn{2}{l}{not reported}\\
  \bottomrule
  \end{tabular}
\end{table}

\begin{table}[ht]
\caption{Contrasts between the effects of the treatments in model \eqref{eq:fixed_eff}\label{tab:marg_eff}. \emph{P-values} are corrected for multiple testing.}
\centering
\begin{tabular}{cccccc}
  \toprule
\textbf{Contrast} & \textbf{Estimate} &  \textbf{Std. Err.} &  \textbf{z-value} & \textbf{Pr($>|z|$)} \\ 
  \hline
\noalign{\vskip 6pt}
   T0 - T1 & 0.040  & 0.070 & 0.563 & 1.080 \\ 
   T0 - T2 & -0.005 & 0.070 & -0.070 & 1.000 \\ 
   T0 - T3 & -0.007 & 0.070 & -0.106 & 1.000 \\ 
   T0 - T4 & -0.010 & 0.070 & -0.141 & 1.000 \\ 
   T1 - T2 & -0.045 & 0.070 & -0.634 & 1.070 \\ 
   T1 - T3 & -0.047 & 0.070 & -0.669 & 1.063 \\ 
   T1 - T4 & -0.050 & 0.070 & -0.704 & 1.056 \\ 
   T2 - T3 & -0.002 & 0.070 & -0.035 & 1.000 \\ 
   T2 - T4 & -0.005 & 0.070 & -0.070 & 1.000 \\ 
   T3 - T4 & -0.002 & 0.070 & -0.035 & 1.000 \\
   \bottomrule
\end{tabular}
\end{table}

Table \ref{tab:marg_eff} reports pairwise contrasts between treatment effects. \emph{P-values} are computed using Tukey’s correction for multiple testing. As expected, no contrast is statistically significant.

\subsection{Verbal Monetary Incentives: Accuracy Heatmaps by Treatment}
\label{app:heatmaps}

\begin{figure}[ht]
    \centering
    \includegraphics[width=1.0\textwidth]{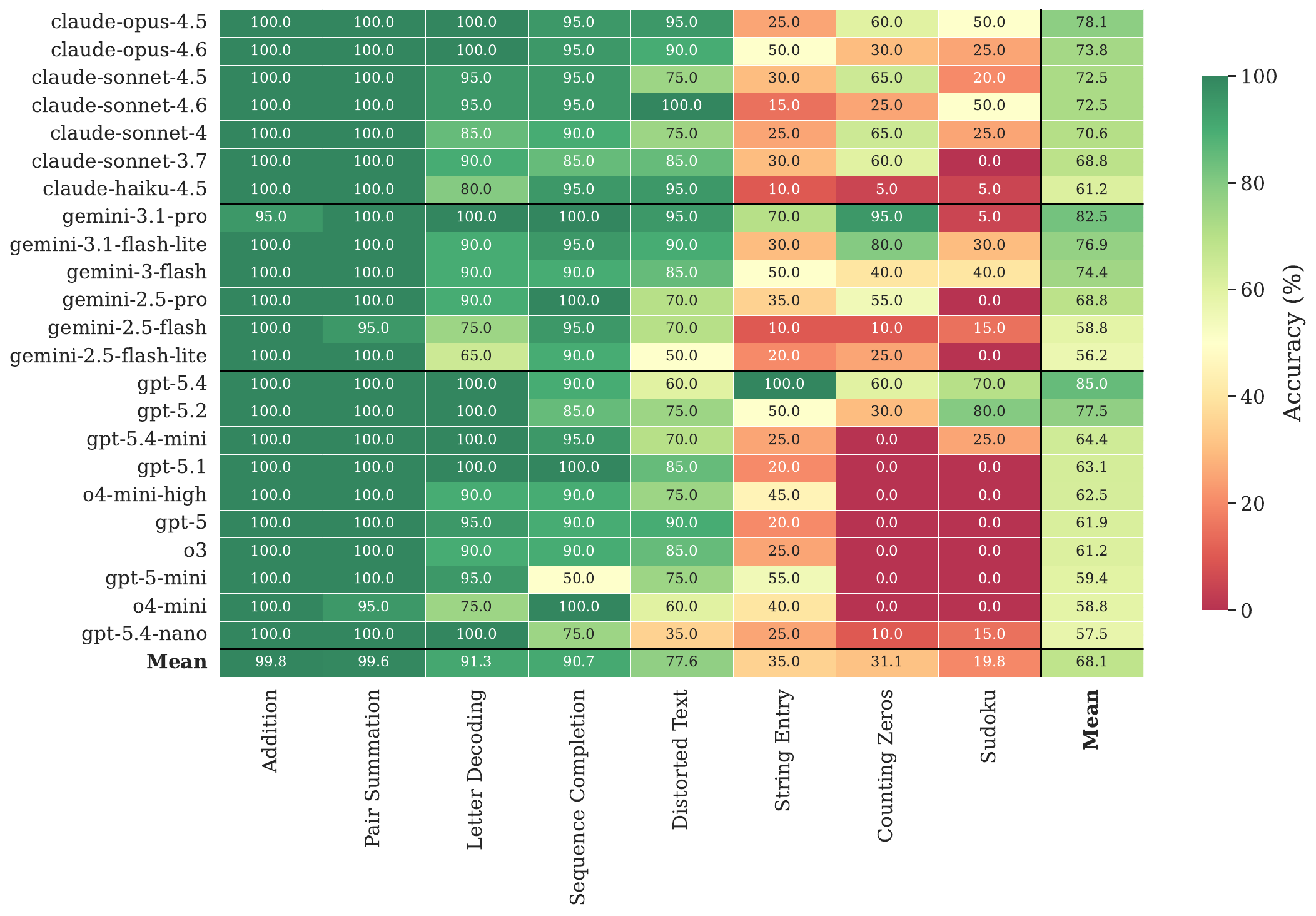}
    \caption{Accuracy (\%) per model and task---T1: Standard, No Incentive.}
    \label{fig:heatmap_t1}
\end{figure}

\begin{figure}[ht]
    \centering
    \includegraphics[width=1.0\textwidth]{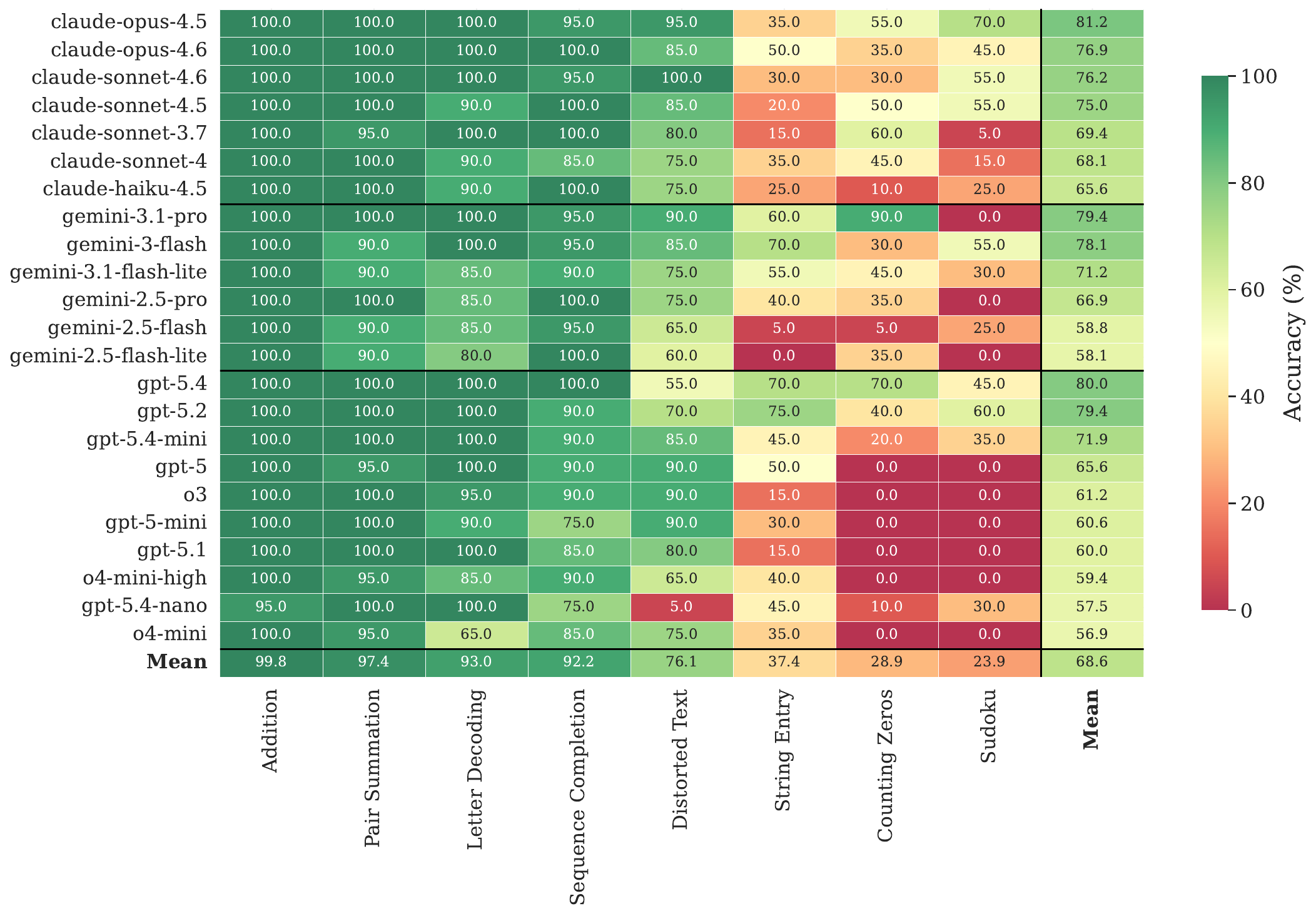}
    \caption{Accuracy (\%) per model and task---T2: Standard, Incentive.}
    \label{fig:heatmap_t2}
\end{figure}

\begin{figure}[ht]
    \centering
    \includegraphics[width=1.0\textwidth]{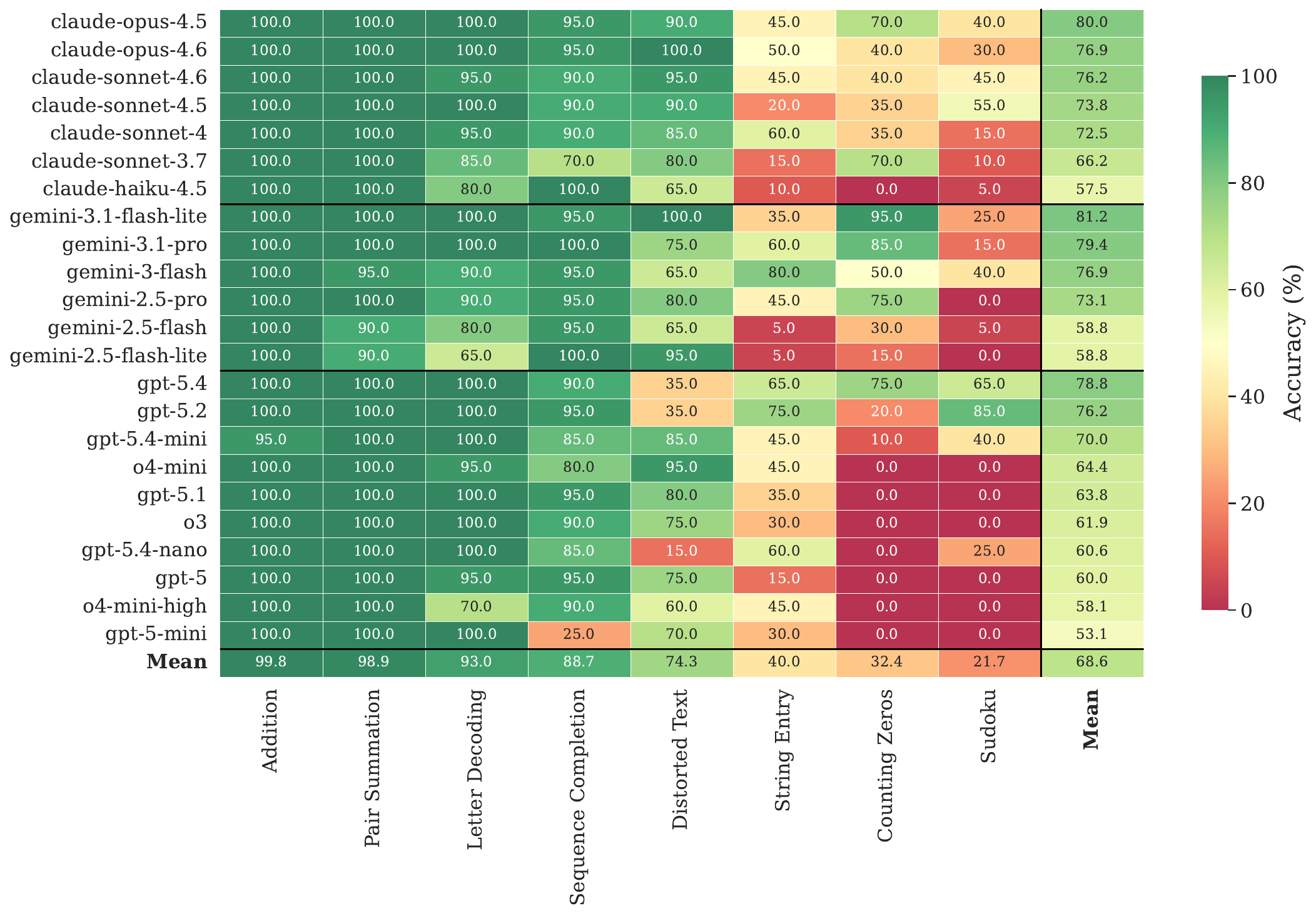}
    \caption{Accuracy (\%) per model and task---T3: Human, No Incentive.}
    \label{fig:heatmap_t3}
\end{figure}

\begin{figure}[ht]
    \centering
    \includegraphics[width=1.0\textwidth]{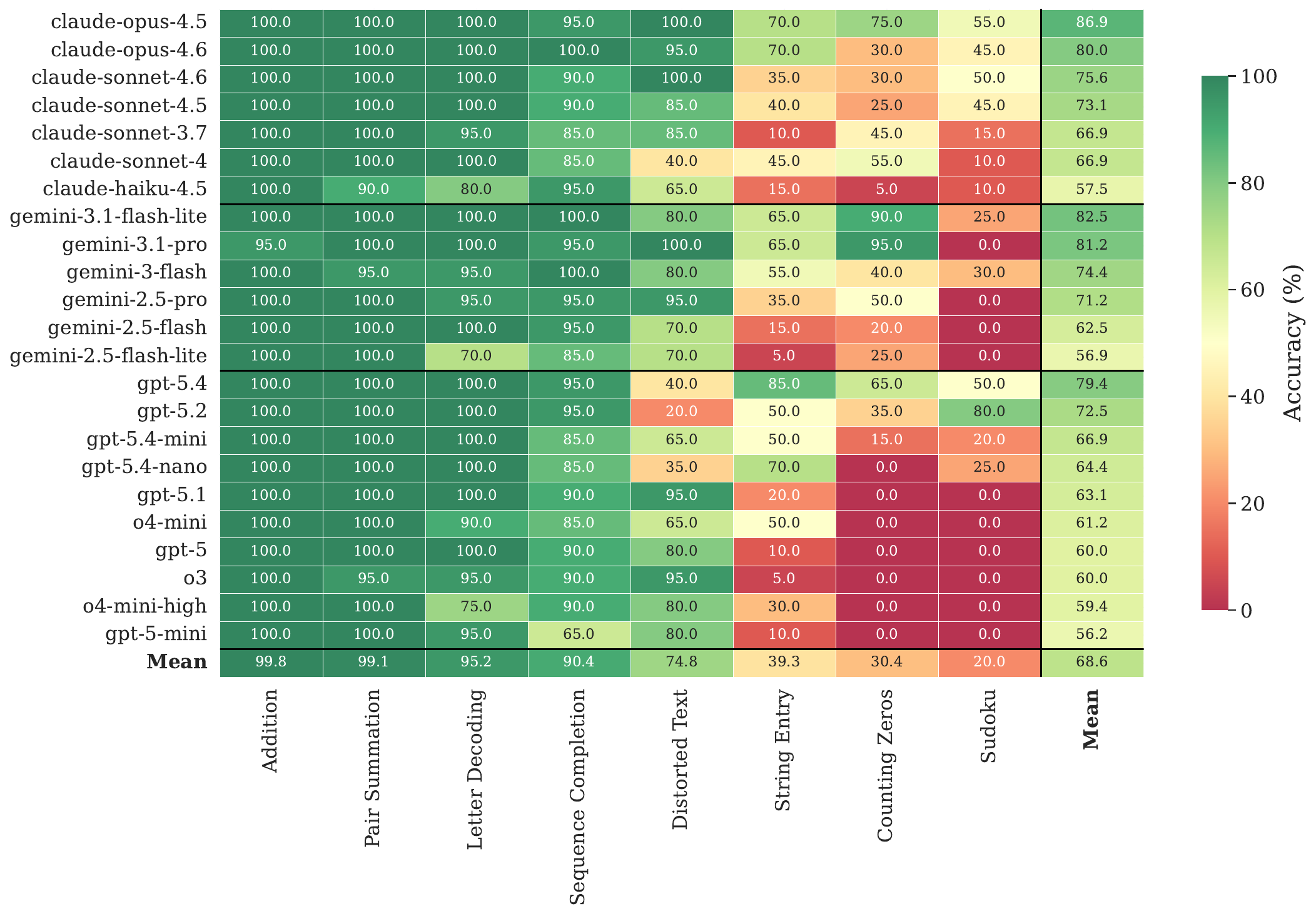}
    \caption{Accuracy (\%) per model and task---T4: Human, Incentive.}
    \label{fig:heatmap_t4}
\end{figure}

Figures~\ref{fig:heatmap_t1}, \ref{fig:heatmap_t2}, \ref{fig:heatmap_t3}, and~\ref{fig:heatmap_t4} complement the main-text heatmap contained in Figure~\ref{fig:heatmap} (which only reports performance under the control treatment T0) with the full accuracy heatmaps (model $\times$ task) for each of the four treatments T1 to T4.

\vspace{-1em}

\vspace{-1em}

\clearpage

\clearpage

\clearpage

\section{Further technical details}

\subsection{Details of the tested LLMs}\label{app:LLMs}

Table~\ref{tab:LLMs} reports the provider, model name, release date, and cost (relative to input and output tokens) of each LLM we tested.

\begin{table}[h]
\caption{Tested LLMs, along with their release dates and prices (USD per million tokens) for input and output, as reported by the OpenRouter API as of March 23, 2026.}
\label{tab:LLMs}
\begin{tabular}{@{}llcrr@{}}
\toprule
\textbf{Provider}  & \textbf{Model}            & \textbf{Released} & \textbf{In(\$/Mt)} & \textbf{Out(\$/Mt)} \\ \midrule
\textbf{Anthropic} & Claude4.6-Opus             & 2026-02           & 5.00               & 25.00               \\
                   & Claude4.6-Sonnet           & 2026-02           & 3.00               & 15.00               \\
                   & Claude4.5-Opus             & 2025-11           & 5.00               & 25.00               \\
                   & Claude4.5-Sonnet           & 2025-09           & 3.00               & 15.00               \\
                   & Claude4.5-Haiku            & 2025-10           & 1.00               & 5.00                \\
                   & Claude-4-Sonnet             & 2025-05           & 3.00               & 15.00               \\
                   & Claude-3.7-Sonnet           & 2025-02           & 3.00               & 15.00               \\ \midrule
\textbf{Google}    & Gemini-3.1-Pro       & 2026-02           & 2.00               & 12.00               \\
                   & Gemini-3.1-Flash Lite & 2026-03           & 0.25               & 1.50                \\
                   & Gemini-3-Flash       & 2025-12           & 0.50               & 3.00                \\
                   & Gemini-2.5-Pro              & 2025-06           & 1.25               & 10.00               \\
                   & Gemini-2.5-Flash            & 2025-06           & 0.30               & 2.50                \\
                   & Gemini-2.5-Flash Lite        & 2025-07           & 0.10               & 0.40                \\ \midrule
\textbf{OpenAI}    & GPT-5.4                   & 2026-03           & 2.50               & 15.00               \\
                   & GPT-5.4-Mini               & 2026-03           & 0.75               & 4.50                \\
                   & GPT-5.4-Nano               & 2026-03           & 0.20               & 1.25                \\
                   & GPT-5.2                   & 2025-12           & 1.75               & 14.00               \\
                   & GPT-5.1                   & 2025-11           & 1.25               & 10.00               \\
                   & GPT-5                     & 2025-08           & 1.25               & 10.00               \\
                   & GPT-5-Mini                 & 2025-08           & 0.25               & 2.00                \\
                   & o3                        & 2025-04           & 2.00               & 8.00                \\
                   & o4-mini                   & 2025-04           & 1.10               & 4.40                \\
                   & o4-mini High       & 2025-04           & 1.10               & 4.40                \\ \bottomrule
\end{tabular}
\end{table}

\subsection{API Cost by Model}
\label{app:cost}

\begin{figure}[ht]
    \centering
    \includegraphics[width=\textwidth]{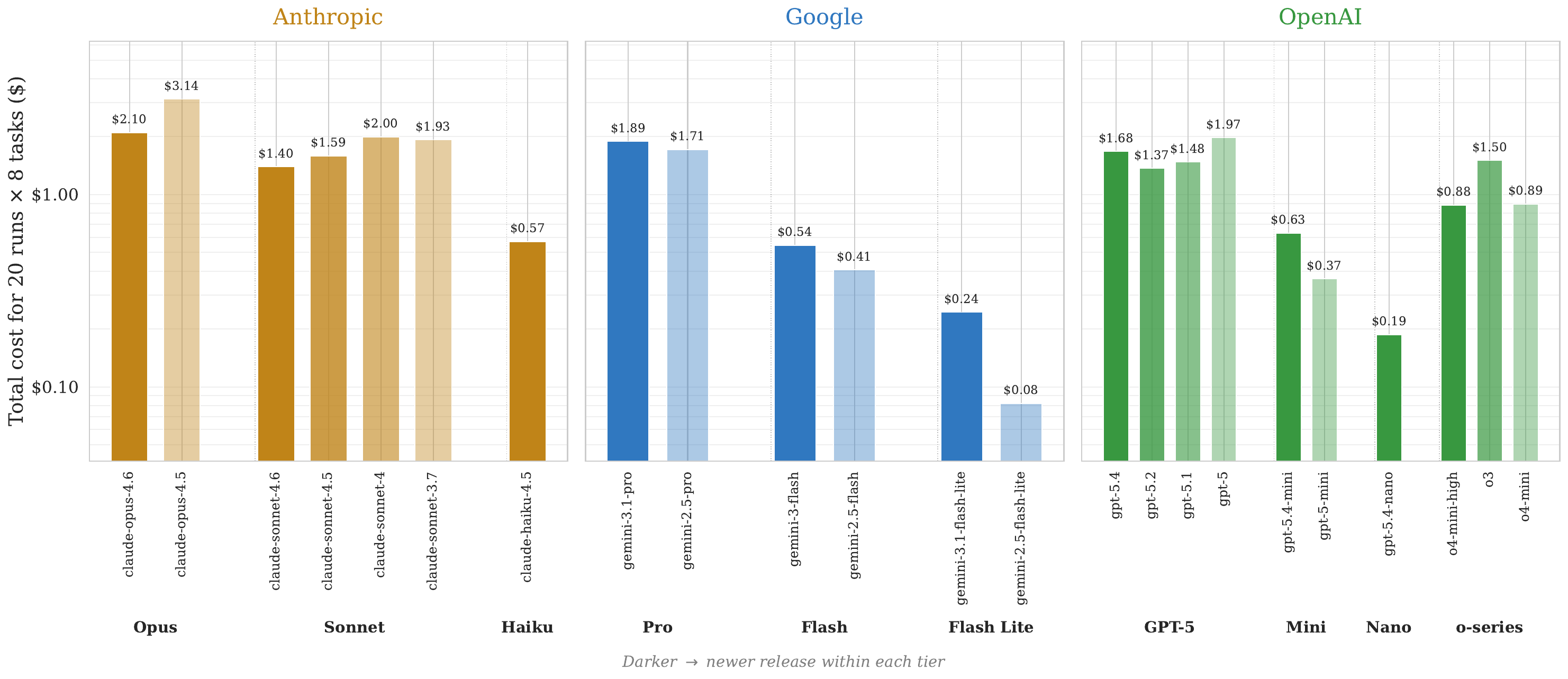}
    \caption{Average API cost per task (\$) $\bar{C}_m$ by model under the control treatment (T0), computed as in the text above. Models are grouped by provider and tier; darker shades indicate newer releases within the same tier. Error bars denote standard errors.}
    \label{fig:cost_per_model}
\end{figure}

Figure~\ref{fig:cost_per_model} reports the average API cost per task for each model under the control treatment (T0). For completeness, we restate here the cost definition introduced in Section~\ref{sec:cost-accuracy}. The total cost for model $m$ on task $t$ is \[C_{m,t} := \sum_{i=1}^{N} \left( p_m^{\text{in}} \cdot T_{m,t,i}^{\text{in}} + p_m^{\text{out}} \cdot T_{m,t,i}^{\text{out}} \right),\] where $N$ is the number of runs, $T_{m,t,i}^{\text{in}}$ and $T_{m,t,i}^{\text{out}}$ are the input (textual prompt plus the image representing the task instance) and output (reasoning plus answer) tokens for model $m$ on the $i$-th run of task $t$, and $p_m^{\text{in}}$, $p_m^{\text{out}}$ are the per-token input and output prices for model $m$. The average cost per model is then \[\bar{C}_m := \frac{1}{|\mathcal{T}|} \sum_{t \in \mathcal{T}} C_{m,t},\] where $\mathcal{T}$ is the set of tasks.

\end{document}